\documentstyle[prl,aps,floats,epsf]{revtex}
\font\msbm=msbm9  scaled \magstep 1    
\def\kappa{\hbox{\msbm\char"7B}}       
\begin{document}
\draft
\tighten
\twocolumn[\hsize\textwidth\columnwidth\hsize\csname
@twocolumnfalse\endcsname
\title{ Transitions of I- and II-order in magnetic field for superconducting \\
cylinder from self-consistent solution of Ginzburg--Landau equations}
\author{G.F. Zharkov}
\address{P.N. Lebedev Physics Institute, Russian Academy of Sciences,
 Moscow, 117924, Russia}
\date{October 2, 2000}
\maketitle
\begin{abstract}
Basing on self-consistent solution of non-linear GL-equations, the phase
boundary is found, which divides the regions of I- and II-order phase
transitions  of a superconducting cylinder in magnetic field to normal state.
This boundary is a complicated function of the parameters ($m,\, R,\, \kappa$)
($m$ is the vorticity, $R$ is the cylinder radius, $\kappa$ is the GL-parameter),
which does not coincide with the simple phase boundary $\kappa=1/\sqrt{2}$,
dividing the regions of I- and II-order phase transitions in infinite (open)
superconducting systems.

\end{abstract}
\pacs{ }
\vskip 2pc ] 
\narrowtext

\section{INTRODUCTION}
The GL-theory is widely used for studying the general properties of the
superconducting state. This theory leads to two coupled non-linear equations
for the order parameter $\psi$ and the magnetic field vector-potential
${\bf A}$, which are usually solved using various simplifying assumptions. In
a number of papers [2--8] the particular case was considered of a long
superconducting cylinder of radius $R$, placed in the axial magnetic field $H$.
In this case the 3-dimensional GL-equations reduce to one-dimensional form,
which enables one to find numerically the exact self-consistent solutions. In
such way it is possible to study specific non-linear effects, as well as the
role of the sample boundary. For instance, it was shown in [6,7], that
one-dimensional solution for the order parameter $\psi$ (with fixed vorticity
$m$ and varying $H$) may change its form either gradually (in one interval of
parameters $(R,\kappa$), $\kappa$ is the GL-parameter), or abruptly (in the
other interval of $(R,\kappa)$), undergoing I-order jump transformation. Such
jump transformations, in principle, may be observable, because they are
accompanied by jumps of the magnetisation, $M(H)$.

In the present paper the phase boundary is found, which divides the region of
parameters ($R,\kappa$), where the superconducting solution (of fixed $m$)
ends up (in the increasing external field) by I-order jump to normal state
($\psi\equiv 0$), from the region ($R,\kappa$), where the solution vanishes
gradually, by II-order phase transition. This phase boundary is a complicated
function of $R$ and $\kappa$, different from the simple boundary
$\kappa=1/\sqrt{2}$, which divides the I- and II-order phase transitions in
infinite (open) superconductors [9]. Other topics are also touched on (such as
metastability, the paramagnetic Meissner effect, the pinning of vortices to the
sample boundary, the linearized equation approximation, etc.).

In Sec. II the problem is formulated and the basic GL-equations, used in
calculations, are written. Sec. II contains the numerical results, alongside
with necessary comments. In Sec. IV the results are summarized and discussed.

\section{EQUATIONS}

Below the case is considered of a long superconducting cylinder of radius $R$,
in the external magnetic field $H\ge 0$, which is parallel to the cylinder
element. In the cylindrical co-ordinates the system of GL-equations may be
written in dimensionless form [6]
$$
{ {d^2U} \over {d\rho^2} } - {1\over \rho}{ {dU}\over {d\rho} } -
\psi^2 U=0,                                                    \eqno(1)
$$
\vskip -0.5cm
$$ { {d^2 \psi} \over {d\rho^2} } + {1\over\rho}
{{d\psi}\over{d\rho}} + \kappa^2 (\psi - \psi^3) - { {U^2 } \over {\rho^2}
}\psi =0.                                                     \eqno(2)
$$
Here $U(\rho)$ is the dimensionless field potential; $b(\rho)$ is the
dimensionless magnetic field; $\psi(\rho)$ is the normalized order parameter;
$\rho=r/\lambda$, $\lambda$ is the field penetration length;
$\lambda=\kappa\xi$, where $\xi$ is the coherence length, $\kappa$ is the
GL-parameter. The dimensioned potential $A$, field $B$ and current $j_s$ are
related to the corresponding dimensionless quantities by the formulae [6]:
$$ A={ {\phi_0} \over {2\pi\lambda} }{ U + m \over \rho },\quad
B={ {\phi_0} \over {2\pi\lambda^2} }b,\quad
b={1\over \rho}{{dU}\over{d\rho}},
$$
\vskip -0.5cm
$$
j(\rho)=j_s\Big/ { {c\phi_0} \over {8\pi^2\lambda^3} }=
-\psi^2 {U\over \rho},\quad  \rho = {r\over \lambda}.              \eqno(3)
$$
(The field $B$ in (3) is normalized by $H_\lambda=\phi_0/(2\pi\lambda^2)$,
with $b=B/H_\lambda$; instead of $H_\lambda$ one can normalize by
$H_\xi=\phi_0/(2\pi\xi^2)$, or by
$H_{\kappa} = \phi_0/(2\pi\xi\lambda) = H_\xi/\kappa$. The coefficients in (1),
(2) would change accordingly.) The vorticity $m$ in (3) specifies how many flux
quanta are associated with the vortex, centered at the cylinder axis (the
so-called giant-vortex state).

The boundary conditions to Eq. (1) are [7]:

$$ U\big|_{\rho =0} = -m,\quad
 \left. { {dU}\over{d\rho} }\right|_{\rho =\rho_1}=h_\lambda.      \eqno(4)
$$
where $\rho_1=R/\lambda$, $h_\lambda=H/H_\lambda$.

The boundary conditions to Eq. (2) are [7]:
$$
\left. {d\psi \over d\rho} \right|_{\rho =0} =0, \quad
\left. {d\psi \over d\rho} \right|_{\rho=\rho_1} =0 \quad (m=0),
$$
\vskip-0.9cm
$$
                        \eqno(5)
$$
\vskip-0.9cm
$$
\psi|_{\rho=0}=0,\quad
\left. { d\psi \over d\rho} \right|_{\rho=\rho_1} =0 \quad (m>0).
$$

The magnetic moment (or, magnetisation) of the cylinder, related to the unity
volume, may be written in a form
$$
{M\over V}={1\over V}\int  {B-H \over 4\pi }dv = { B_{av}-H \over 4\pi },
$$
$$
B_{av}={1\over V}\int B({\bf r})dv={1\over S}\Phi_1,
$$
where $B_{av}$ is the mean field value inside the superconductor, $\Phi_1$ is
the total magnetic flux, confined in the cylinder. In the normalization (3),
denoting $\overline{b}=B_{av}/H_\lambda$, $h_\lambda=H/H_\lambda$,
$M_\lambda=M/H_\lambda$, one finds
\begin{eqnarray*}
\qquad\qquad\quad 4\pi M_\lambda=\overline{b}-h_\lambda, \quad
\overline{b}={2\over\rho_1^2}(U_1+m), \quad \quad   (6)  \\
\phi_1={\Phi_1 \over \phi_0}=U_1+m, \quad
U_1=U(\rho_1),\quad \rho_1={R\over\lambda}.
\end{eqnarray*}

Accordingly, the normalized Gibbs free energy of the system may be written as
[7]
$$
\Delta g=\Delta G\Big/ \left( { H_{c{\rm m}}^2 \over 8\pi } V \right)=
g_0-{8\pi M_\lambda \over \kappa^2} h_\lambda +{4m \over \kappa^2}
{b(0)-h_\lambda \over \rho_1^2},                          \eqno(7)
$$
\vskip -0.7cm
$$
g_0={2\over\rho_1^2} \int_0^{\rho_1}
\rho d\rho \left[ \psi^4-2\psi^2+{1\over\kappa^2} \left(
{d\psi \over d\rho} \right)^2 \right].
$$
Here $\Delta G=G_s-G_n$ is the difference of free energies in superconducting
and normal states; $b(0)=B(0)/H_\lambda$, $B(0)$ is the magnetic field at the
cylinder axis; $H_{c{\rm m}}=\phi_0/(2\pi\sqrt{2}\lambda\xi)$ is thermodinamical
critical field of massive superconductor; $g_0$ is the condensation energy with
account for the order parameter gradient. The expressions (6), (7) may be used
for calculating the corresponding quantities. [Instead of $\rho_1$ the notation
$R_\lambda=R/\lambda\equiv\rho_1$ will be used below.]

\section{NUMERICAL RESULTS}

The solutions of Eqs. (1)--(5) depend on the space co-ordinate $\rho$ and several
parameters, for instance, $\psi(\rho)=\psi(m,R_\lambda,\kappa,h_\lambda;\rho)$
(analogously for the potential $U(\rho)$ and the field $b(\rho)$). Let the
vorticity $m$ be fixed ($m=0,\,1,\,2,\dots$) and consider at first the case
$m=0$ (the vortex-free Meissner state). Consider the plane of parameters
($R_\lambda,\kappa$) (see Fig. 1($a$)). In every point of this plane there
exists a set of solutions of Eqs. (1)--(5), which depend parametrically on
the external field $h_\lambda$. (Several points, laying along the line
$R_\lambda=4$ are numerated as {\it 1-6}.) One may envisage a peep-hole, pearced
in every point ($R_\lambda,\kappa$), which allows to see the content of the
corresponding sub-volume. The set of solutions $\psi(h_\lambda;\rho)$ is unique
for each sub-volume and may be characterised, for instance, by the field
dependence of the maximal value of the order parameter $\psi_{max}(h_\lambda)$,
or by the form of the magnetisation curve $M_\lambda(h_\lambda)$ (6). The
examples of such dependencies in different points of the plane
$(R_\lambda,\kappa)$ are given in Figs. 1($b,c$) (only the case
$h_\lambda\ge 0$ is considered; some illustrations for the case $h_\lambda<0$,
as well as the corresponding co-ordinate dependencies, may be found in [6,7]).

It is clear from Fig. 1($b$), that the characteristic behavior
$\psi_{max}(h_\lambda)$ depends essentially on the value of $\kappa$. For small
$\kappa$, the value $\psi_{max}(h_\lambda)$ terminates (the curves {\it 1-4})
by jump at some point $h_\lambda=h_s$, where (if $h_\lambda$ is increased)
the I-order phase transition to normal state ($\psi(\rho)\equiv 0$) occures.
The region, where the superconducting solutions terminate by I-order phase
transition, is marked in Fig. 1($a$) as $s_{\rm I}$.

For larger $\kappa$ (the curves {\it 5,6}) there is also a jump in
$\psi_{max}(h_\lambda)$ at some point $h_\lambda=h_s$, but with a "tail"
appearing on the curve. If the field $h_\lambda$ increases, the superconducting
solutions {\it 5,6} vanish gradually at the point $h_c$, by II-order phase
transition to normal state. The region, where the superconducting solutions
terminate by II-order phase transition, is marked in Fig. 1($a$) as
$s_{\rm II}$. [The appearance of the tail on the magnetisation curve means the
transition of the solution to the edge-suppressed form, see [7] for details.]

It is evident, that for small radius cylinder ($R_\lambda<1.69$) the
superconducting solution terminates by II-order phase transition, even in type-1
(i.e. small $\kappa$) superconductors [10]. The transformation of the solutions
with diminishing radius $R_\lambda$ is illustrated in Fig. 2 for $\kappa=0.7$.

Note, that if the line $\kappa=1$ in Fig. 1($a$) is followed from large to small
$R_\lambda$, the superconducting states, which lay along this line, display at
first the II-order phase transition in magnetic field (for larger $R_\lambda$),
then the I-order (for intermediate $R_\lambda$), and again the II-order (for
smaller $R_\lambda$). Only, if $\kappa>1.05$, all the solutions display
II-order behavior.

Notice also, that the state $m=0$ is totally diamagnetic ($-4\pi M_\lambda>0$).

Because in every point of $s_{\rm II}$-region the order parameter vanishes by
II-order phase transition ($\psi_{max}\to 0$, see Fig. 1($b$)), the
superconducting phase boundary in magnetic field, $h_c$, may be found
analitically, by linearizing the system (1), (2) (with account, that
$\psi\ll 1$ and $b\approx h_\lambda$), and passing to single linear equation
for the order parameter [11] , whose solution may be expressed in terms of the
Kummer functions (see also [3,4,5,12]). However, inside $s_{\rm I}$-region
(where the solution terminates by jump from a finite value $\psi_{max}$ to zero)
the phase boundary $h_s$ (i.e. the highest field $h_\lambda$ still compatible
with the superconductivity) can not be found by solving the linearized equation,
but full system (1)--(5) is needed.

The analogous investigation can be carried out in the case $m=1$ (see Fig. 3),
with a single vortex on the cylinder axis.

In Fig. 3($a$) are shown: the region $s_{\rm I}$, where the superconducting
state terminates (if the field is increased) by I-order jump to the normal
state, having finite value $\psi_{max}$ at the transition point; the region
$s_{\rm II}$, where the superconductivity vanishes by II-order phase
transition; the curve $s_{\rm I-II}$, which represents the boundary between
I- and II-order phase transitions.

The behavior of the order parameter $\psi_{max}(h_\lambda)$ and
of the magnetisation $M_\lambda(h_\lambda)$ in different points of the plane
$(R_\lambda,\kappa)$ are shown in Figs. 3($b,c$) (and in Fig. 4). For small
$\kappa$ (the curves {\it 1,2}) the solutions terminate by I-order jump. When
the line $s_{\rm I-II}$ is crossed, the tail appears on the curves {\it 3,4},
which widens, if $R_\lambda$ and $\kappa$ increase. If $R_\lambda$ diminishes
(Fig. 4), the magnitude of the jump in $\psi_{max}$ also diminishes, and the
solutions terminate (if the field is increased) by II-order phase transition
to normal state.

On the curve $C_{ns}$ (Fig. 3($a$)) the value $\psi_{max}=0$. The letter $n$
denotes the normal metal region ($\psi\equiv 0$); here the superconducting
state ($m=1$) is impossible. [In this region the radius $R_\lambda$ is too
small, and the vortex own field is too strong to be confined within the
mesoscopic sample.]\, It is evident, that when the radius $R_\lambda$
diminishes (but $\kappa$ is fixed) the transition from $s$- to $n$-state always
is II-order phase transition, however the width of the region between the
curves $s_{\rm I-II}$ and $C_{ns}$ (where II-order transition exists) gets
very small for small $\kappa$. The curve $C_{ns}$ may be well approximated by
the dependence $R_\lambda\sim a/\kappa$ (or $R_\xi=\kappa R_\lambda=a$), with
$a=1.34$.

Notice, that in any point of $s$-region in Fig. 3($a$) the magnetisation
function $M_\lambda(h_\lambda)$ (Fig. 3($c$)) has two parts: the paramagnetic
($M_\lambda>0$) and diamagnetic ($M_\lambda<0$). This is because the
superconducting current has two components, $j_s=j_p+j_d$. One of these currents
($j_p$) screens the own field of the vortex ($m=1$) and flows around the vortex
axis in counter-clockwise direction (the paramagnetic current). The second
current ($j_d$) screens out the external field $h_\lambda$ and flows near the
cylinder surface in clock-wise direction (the diamagnetic current). Depending
on which of these currents prevail, the magnetisation (or, equivalently, the
magnetic moment $M_\lambda=(1/2c)\int [{\bf j}_s{\bf r}]dv$) can change sign,
as function of $h_\lambda$ (see [8] for details). Recall, that in the
vortex-free state ($m=0$) there exists only diamagnetic current, i.e.
$M_\lambda<0$, see Fig. 1($c$). In presence of the vortex ($m=1$), but in
absensce of the external field $(h_\lambda=0)$, the screening current and the
magnetic moment correspond to the paramagnetic state. This state is metastable,
because the vortex-free state posesses smaller free-energy, than the state
$m=1$ [2--8].

The curve $P_0$ in Fig. 3($a$) corresponds to the minimal radius $R_\lambda$,
when the paramagnetic vortex state ($m=1$) can still exist inside the
homogenious cylinder in absence of the field ($h_\lambda=0$). [The metastable
vortex is held inside by the pinning to the cylinder boundary.]\, In those
points $(R_\lambda,\kappa)$, which lay below the curve $P_0$, to held the
vortex inside the cylinder it is necessary to impose a finite external field,
$h_\lambda>0$. (This corresponds to the field stimulated and re-entrant
superconductivity [2--8].)\, Notice, that if the cylinder radius $R$ and the
parameter $\kappa$ are fixed, to cross the paramagnetic pinning boundary
($P_0$) it is sufficient to vary only the sample temperature, because
$R_\lambda=R/\lambda(T)$.

The presence of a smooth tail in the function $\psi_{max}(h_\lambda)$
[Figs. 3($b$) and 4($a$)] allows (as in the case $m=0$) to use the linear
approximation ($\psi\ll 1$) for finding the upper boundary of the
superconducting state, $h_c$. In the region of I-order jumps [$s_{\rm I}$ in
Fig. 3($a$)], where the function $\psi(\rho)$ is not small, the linear approach
fails and more rigorous analysis, based on full system of non-linear equations
(1)--(5), is neccessary. [The boundaries $s_{\rm I-II}(\kappa)$ and
$C_{ns}(\kappa)$ themselves can not be found from the linear equation, because
the latter does not depend on $\kappa$ [11]. The comparison of the results of
the rigorous and linear analysis will be reported elsewhere.]

Similarly, one can consider the higher giant-vortex states ($m>1$, see Fig. 5
for $m=2$). Here also exist the boundaries of I- and II-order phase transitions,
the jumps on the magnetisation curves, the paramagnetic and diamagnetic currents,
and other peculiarities, which are analogous to those presented in Figs. 1--4.

\section{Conclusion and discussion}

Basing on self-consistent solution of non-linear system of GL-equations, the
boundary, $s_{\rm I-II}$, is found, which separates the regions, where the
superconducting state of the cylinder is destroyed by the external magnetic
field either by I-order jump (the region $s_{\rm I}$), or gradually, by II-order
phase transition (the regioin $s_{\rm II}$). This boundary is a complicated
function of the parameters ($m,R_\lambda,\kappa$) [see Figs. 1($a$), 3($a$), 5].

Note, that in the case of an infinite (open) superconductor the phase
boundary between I- and II-order transitions lays at the value
$\kappa=1/\sqrt{2}$ [9]. (At this value the surface energy at the interface
of superconducting and normal metals vanishes, and the magnetisation $M(H)$
acquires a smooth tail [9].)\, However, the case of infinite superconductor is
degenerated, in the sense that the total number of vortices in the open system
can not be defined. Due to this degeneration there are many solutions of the
system (1), (2) with different $m$, and it is possible to consider the
superconducting state as a linear combination of states with different
vorticities $m$ [9]. In the bounded system this degeneracy is removed, and it
is necessary to consider the states of fixed vorticity $m$ (the quantum number
$m$ is now a topological invariant). [It is easy to prove that self-consistent
solution of Eqs. (1)--(5) with $\psi\ne 0$ is unique for every value of
$h_\lambda$.]\, The mentiond difference of the $s_{\rm I-II}$-boundary from
the value $\kappa=1/\sqrt{2}$ is due to the difference in geometries and to the
account of the space-quantization effects, present in the bounded system.
[To trace the limiting transition from the bounded to open geometry, it might
be neccessary to consider the case of flattend elliptical cylinder, which
models the geometry of an infinite slab, adopted in [9].]

Mention in conclusion, that the main attention in the present work was paid to
the mathematical side of the problem: to describe the $s_{\rm I-II}$-boundary
basing on formal solutions of GL-equations. The important physical question of
compairing the Gibbs free energies of various mathematically possible states,
and finding the most stable ground state (which the system would occupy in
equilibrium), was put aside. (Some illustrations of the Gibbs free energy
behavior, found from Eq. (7), are given, for instance, in [6-8].)\, In
justification, it may be reminded, that the physical system may occupy not only
the groung state, but also the excited metastable states of higher energy. [In
particular, the controversial paramagnetic Meissner effect may be attributed to
the metastable vortex states in the mesoscopic system, see [8] for details).
The formal solutions of GL-equations describe all possible states, including
stable and metastable one, thus the full analysis, based on these equations,
may be pertinent to the experiment. However, the case of infinitely long
cylinder, considered above, approximates rather poorly the geometry, used in
real experiments. In this respect, the superconducting disks, considered in
[4,5], are more adequate, though it would be more difficult to obtain rigorous
solutions for the bounded 3-dimensional sample. Thus, further analysis of the
questions, touched on in the present paper (as well as possible connection with
experiment), is necessary.

\section{Acknowledgments}

I am grateful to V. L. Ginzburg for the interest in this work and valuable
discussions. I thank also F. M. Peeters and J. J. Palacios for sending the
reprints of recent papers, where closely connected problems are considered.

\centerline{\bf Figures captions}

Fig. 1. ($a$) -- The boundary ($s_{\rm I-II}$) between regions ($s_{\rm I}$ and
$s_{\rm II}$), where I- or II-order phase transitions to normal state
($\psi\equiv 0$) in magnetic field occure.
($b$) -- The behavior $\psi_{max}(h_\lambda)$ in the points {\it 1-6}
($m=0,\,R_\lambda=4$) in Fig. 1($a$). In the region $s_{\rm I}$ the order
parameter vanishes by I-order jump. In the region $s_{\rm II}$ the order
parameter $\psi_{max}(h_\lambda)$ has a "tail", and vanishes smoothly, by
II-order phase transition.
($c$) -- Analogous behavior for magnetisation, $M_\lambda(h_\lambda)$.
The peep-holes {\it 1-9} in ($a$) are pearced in the points:
{\it 1} -- $\kappa=0.2$, {\it 2} -- $\kappa=0.4$, {\it 3} -- $\kappa=0.7$,
{\it 4} -- $\kappa=1$, {\it 5} -- $\kappa=1.0.5$, {\it 6} -- $\kappa=1.2$
($R_\lambda=4$); {\it 7} -- $R_\lambda=3$, {\it 8} -- $R\lambda=2$,
{\it 9} -- $R\lambda=1.5$ ($\kappa=0.7$).

Fig. 2. The dependencies: ($a$) -- $\psi_{max}(h_\lambda)$ and ($b$) --
$M_\lambda(h_\lambda)$ for $m=0,\,\kappa=0.7$. The numeration of curves
correspond to points {\it 3,7--9} in Fig. 1($a$).

Fig. 3. Analogous to Fig. 1, but for $m=1$. The dashed curve $C_{ns}$ in
Fig. 3($a$) separates the normal ($n$) and superconducting ($s$) regions.
The curve $P_0$ marks the points ($R_\lambda,\kappa$), where the metastable
vortex state ($m=1$) may still exist in absence of the field ($h_\lambda=0$)
due to the pinning to the boundary. Below the curve $P_0$ the vortex state
may exist only in presence of finite external field ($h_\lambda>0$, see the
curves {\it 1,2} in Figs. 3($b,c$)). (This is an example of the field
stimulation effect, or re-entrant superconductivity.) The peep-holes {\it 1-8}
in ($a$) are pearced in the points: {\it 1} -- $\kappa=0.35$,
{\it 2} -- $\kappa=0.4$, {\it 3} -- $\kappa=0.7$, {\it 4} -- $\kappa=1$,
{\it 5} -- $\kappa=1.07$, {\it 6} -- $\kappa=1.2$ ($R_\lambda=4$);
{\it 7} -- $R_\lambda=3$, {\it 8} -- $R_\lambda=2.4$ ($\kappa=0.7$).

Fig. 4. Analogous to Fig 2, but for $m=1$. The presence of paramagnetic
($M_\lambda>0$) and diamagnetic ($M_\lambda<0$) parts of magnetisation is
evident in Fig. 4($b$).

Fig. 5. Analogous to Figs. 1($a$) and 3($a$), but for $m=2$. The vertical
asymptote $\kappa=0.94$ is the same for $m=0,1,2$. This is natural, because for
large radii ($R_\lambda\gg 1$) the influence of the vortex field is negligible.
The bottom of the curve $s_{\rm I-II}$ lays at $R_\lambda=2.78$ (with
$R_\lambda=2.45$ for $m=1$, and $R_\lambda=1.69$ for $m=0$). The dashed curve
$C_{ns}$ is well approximated by the dependence $C_{ns}\approx 1.81/\kappa$
(the dotted line).

\end{document}